\documentclass[a4paper,11pt]{article}

\usepackage{amssymb}
\usepackage{amsmath}
\usepackage[dvipdfmx]{graphicx}
\usepackage[breaklinks=true]{hyperref} 
\usepackage[normalem]{ulem} 
\usepackage{color}
\usepackage{ulem}
\usepackage{cite}
\usepackage{appendix}

\begin{document}




\title{
Peccei-Quinn Symmetry Breaking via Asymptotically Safe Dynamical Scalegenesis: 
A Walking Axicolor and Axion} 


\author{Hiroyuki~Ishida$^1$, Shinya~Matsuzaki$^2$, and Xiao-Chang Peng$^{2,3}$}
\date{}
\maketitle
\vspace{-10mm}

\begin{center}
$^1${\small 
  Theory Center, IPNS, KEK, Tsukuba, Ibaraki 305-0801, Japan
}

$^2${\small 
	Center for Theoretical Physics and College of Physics, Jilin University, Changchun, 130012,
China
}

$^3${\small 
Department of Physics, Brown University, Providence, RI 02912, USA
}
\end{center}
%
%
%
%
%
%
%
%


%

\begin{abstract}

Pecci-Quinn (PQ) symmetry breaking by perturbative dynamics 
would suffer from a hierarchy problem, just like 
the electroweak symmetry breaking in the standard model. 
The dynamics of the axion, associated with the PQ symmetry breaking, 
would also involve a triviality problem. 
We provide a paradigm to resolve those two problems potentially existing in the PQ symmetry breaking scenario, 
with keeping successful axion relaxation for the QCD strong CP phase. 
The proposed theory 
includes an axicolor dynamics with 
the axicolored fermions partially gauged by the QCD color,  
and is shown to be governed by 
an asymptotically safe (AS) fixed point: quantum scale invariance is built.   
The AS axicolor is actually a ``walking" dynamics, which  
dynamically breaks a PQ symmetry, a part of 
the chiral symmetry carried by the axicolored fermions.  
The PQ scale generation is then triggered by the nonperturbative 
dimensional transmutation in the ``walking" dynamics. 
A composite axion emerges as the assosiated Nambu-Goldstone boson. 
That is, no hierarchy or triviality problem is present there. 
The composite axion 
can potentially be light due to the characteristic feature of 
the AS axicolor (``walking" axicolor), 
becomes the QCD axion in the anti-Veneziano limit, and gets heavier by the subleading 
correction. 
The composite axion relaxes the QCD theta parameter, involving 
heavier relaxation partners such as axicolored pseudoscalar mesons,  
and the ultraviolet correction to the relaxation mechanism is protected 
by the established (near) scale invariance during the ``walking" regime.

\end{abstract}







\section{Introduction}



Violation of CP symmetry was theoretically predicted and observed in the weak interactions, 
however, not observed in the strong interactions, which is called strong CP problem. 
(See e.g.,~\cite{Peccei:1977ur,Shifman:1979if,Kim:2008hd}  as reviews.)
One of the most attractive solution to this problem is introducing a global symmetry, 
which is often called Peccei-Quinn (PQ) symmetry~\cite{Peccei:1977hh,Weinberg:1977ma,Wilczek:1977pj,Peccei:1977ur} 
$U(1)_{\rm PQ}$, 
and the pseudo Nambu-Goldstone boson associated with 
the spontaneous breaking of 
this global symmetry is called axion. 


The role of the spontaneous breaking of the PQ symmetry can simply be played by a  
new Higgs boson (a PQ breaking scalar), via an assumed double-well type potential, analogously to 
the standard model~\cite{Kim:1979if,Zhitnitsky:1980tq,Dine:1981rt,Shifman:1979if}. 
However, this kind of simple and perturbative scenario inevitably suffers 
from a hierarchy problem or instability for the PQ breaking scale ($f_a$) 
against power corrections arising from the Planck scale,
similarly to the gauge hierarchy problem caused by the Higgs boson in the standard model~\footnote{  
Moreover, 
quantum gravity would be expected to 
break any global symmetry, which could destabilize 
the axion relaxation mechanism 
~\cite{Holman:1992us,Kamionkowski:1992mf,Barr:1992qq,Ghigna:1992iv}. 
Later we will give comments on this issue in terms of the present framework. 
}.

Furthermore, renormalization group evolution for the PQ-breaking scalar sector would give 
Landau poles 
of the quartic coupling for the PQ breaking scalar: namely 
another triviality problem would be caused. 
This problem may still be left even if 
the PQ symmetry is realized as an 
accidental unbroken symmetry arising from a  
gauge symmetry breaking~\cite{Chun:1992bn,BasteroGil:1997vn,Babu:2002ic,Fukuda:2017ylt,Duerr:2017amf,Bonnefoy:2018ibr,Redi:2016esr,Darme:2021cxx,Nakai:2021nyf,Randall:1992ut,DiLuzio:2017tjx,Lillard:2018fdt,Lee:2018yak,Ardu:2020qmo,Yin:2020dfn,Yamada:2021uze}, where the new gauge coupling may also have a Landau pole.

A simple solution to the former problem is to explain  
the origin of PQ symmetry breaking by a strong dynamics, called axicolor~\cite{Kim:1984pt,Choi:1985cb}. 
In this case, the PQ symmetry is realized as a part of 
the global chiral symmetry carried by the axicolored fermions, 
and is spontaneously broken by the dynamical generation of 
the fermion mass and nonzero chiral condensate 
due to the axicolor which gets strong at low energy, just like 
QCD. 
Thus, axion arises as a composite Nambu-Goldstone boson 
associated with the spontaneous chiral breaking,   
and therefore, no hierarchy problem for the PQ breaking scale 
is present there~\footnote{
For similar frameworks breaking the PQ symmetry, realized as 
a part of composite chiral symmetry, by the dimensional transmutation, 
e.g. see~\cite{Harigaya:2015soa,Yamada:2015waa,Yamada:2021uze}. 
}. 
However, the triviality problem is still left for 
other scalars which are necessarily introduced to make 
other Nambu-Goldstone bosons fully massive.

As for the latter triviality problem, 
the author in Ref.~\cite{Salvio:2020prd} has recently found perturbatively asymptotic-free fixed points for all the couplings in an axion model.
However, in such a perturbative approach the hierarchy problem for the PQ breaking scale 
has been unsolved. 

In this paper, we propose a new possibility that 
realizes a composite axion in asymptotically non-free, 
but nontrivially interacting ultraviolet gauge theory, characterized by  
an asymptotically safe (AS) fixed point manifestly keeping the quantum scale invariance. 
This provides a new paradigm to resolve the hierarchy and triviality 
problems for the axion dynamics, without spoiling successful axion 
relaxation for the QCD strong CP problem.

The theory includes the gauge (called axicolor), Yukawa and scalar sectors.   
We work on the ($1/N_F$, $1/N_C$) dual expansion with $N_F/N_C$ fixed and 
$N_F \gg N_C$ (which was dubbed 
anti-Veneziano limit in the literature~\cite{Matsuzaki:2015sya}), 
where $N_{F}$ and $N_C$  
respectively denote the numbers of flavors and colors.
We then see that the axion arises as a composite pseudo-Nambu-Goldstone boson 
associated with the chiral symmetry breaking for $N_F$ fermions 
in the strong-gauge coupling regime 
of the axicolor.  
The spontaneous breaking of the chiral symmetry is governed by 
a strong QED like dynamics with the constant coupling identified 
as the AS fixed point coupling, that is almost scale-invariant, 
walking dynamics.

The composite axion becomes the QCD axion in the anti-Veneziano limit, 
and can be heavier due to the subleading 
nature.  
The phenomenologically viable relaxation mechanism for the strong CP phase 
works at the subleading 
order of the large $N_F$ limit. 
The QCD CP problem can thus be interpreted as the subleading 
phenomenon, and solved 
by a hybrid relaxation mechanism involving a couple of 
(composite) pseudoscalars, like so-called the ultraviolet solution~\cite{Draper:2016fsr,Agrawal:2017ksf,Gaillard:2018xgk,Gherghetta:2020ofz}. 

Several comments and discussions along the presently proposed scenario are provided in the last section (Conclusion and Discussions).

\section{Asymptotically Safe Axicolor}

The axicolor is introduced as the $SU(N_C)$ gauge with the gauge coupling $g$, 
to which $N_F$-fundamental representation fermions couple. 
We vectorlikely gauge the chiral symmetry in part and identify 
the color triplet as three of $N_F$ fermions,  
such that the $N_F$-vector formed by $F$-fermions goes like 
\begin{align}  
F = (\psi^{\alpha}, \psi_1, \cdots, \psi_{n_F})^T 
\label{F}
\end{align}
with $\alpha =1,2,3$ (QCD colors) 
and $n_F=N_F-3$. 
This QCD color embedding is a generalization of the dynamical axion model 
in~\cite{Choi:1985cb}, which corresponds to $n_F=1$. 

We also introduce color-singlet scalar fields  
which take an $n_F \times n_F$ matrix form ($M$)  
transforming   
in a bifundamental way under $U(n_F)_L \times U(n_F)_R$. 
This chiral scalar matrix $M$ can only have the quartic 
potential terms because of the scale symmetry~\footnote{
The null mass term as well as vanishing other dimensionful coupling terms at the Planck scale  
can be ensured by possible 
existence of another AS fixed point over Planckian scale and its associated large anomalous dimensions, which 
enables to tune those to be zero~\cite{Shaposhnikov:2009pv,Wetterich:2016uxm,Eichhorn:2017als,Pawlowski:2018ixd,Wetterich:2019qzx}. 
\label{foot2}
}: 
\begin{align} 
V_M 
= \lambda_1 {\rm tr}[(M^\dag M)^2] + \lambda_2 ({\rm tr}[M^\dag M])^2
\label{VM}
\end{align}  
with the couplings $\lambda_{1,2}$.

The axicolored fermions also couple to the color singlet 
complex scalars 
in an invariant way under the scale symmetry 
and the chiral $U(n_F)_L \times U(n_F)_R$ symmetry: 
\begin{align} 
{\cal L}_y = - y (\bar{\psi}_L M \psi_R + \bar{\psi}_R M^\dag \psi_L) 
, \label{Ly} 
\end{align}  
where $\psi = (\psi_1, \cdots, \psi_{n_F})^T $ corresponding to 
the subset of $F$ in Eq.(\ref{F}). 
The Yukawa coupling $y$ can always be taken to be real, by field redefinition of 
$\psi_{L,R}$.

For later convenience, we define the fine structure constants  
relevant to the anti-Veneziano limit: 
\begin{align} 
\alpha_g 
&\equiv 
\frac{N_C g^2}{(4\pi)^2}\,, 
\qquad 
\alpha_y \equiv  
\frac{N_C y^2}{(4\pi)^2} 
\,, \notag \\ 
\alpha_{\lambda_1} 
&\equiv 
\frac{n_F \lambda_1}{(4\pi)^2}\,, 
\qquad 
\alpha_{\lambda_2} \equiv  
\frac{n_F^2 \lambda_2}{(4\pi)^2} 
\,. \label{alphas}
\end{align} 
All of them are of $O(N_F^0, N_C^0)$ in the large $N_F$ and $N_C$ 
limit.

The Yukawa coupling $y$ is assumed to be tiny in the AS domain 
(e.g., $10^{-90} \lesssim \alpha_y \lesssim 10^{-6}$ for a later reference) 
and gets asymptotically free. 
The QCD coupling is also assumed to reach the 
asymptotically-free fixed point, 
because the gluonic contribution still dominates over colored fermion's 
even including extra $\psi^\alpha$ in Eq.(\ref{F}). 
Since the Yukawa coupling is tiny, the scalar sector will be almost 
completely external to the axicolor gauge dynamics, so will also become  
irrelevant to the dynamics of emergent 
composite axion itself.



Of particular interest is in the strongly coupled regime for the axicolor, 
which would  
give rise to the spontaneous-chiral symmetry breaking. 
Then the axicolor dynamics responsible for
the chiral symmetry breaking can 
be viewed like a strong QED with the constant gauge coupling. 
In that case, for the dynamical-chiral symmetry breaking,  
we may apply the result 
based on the ladder Schwinder-Dyson gap equation, 
which has extensively been studied in several contexts~\cite{Johnson:1964da,Johnson:1967pk,Maskawa:1974vs,Maskawa:1975hx,Fukuda:1976zb,Miransky:1984ef,Yamawaki:1985zg,Bando:1986bg,Aoki:1989su} (see also a review~\cite{Yamawaki:1996vr}.). 
In terms of the current coupling convention in Eq.(\ref{alphas})
the critical coupling and the generated fermion dynamical mass ($m_F$), following 
so-called the Miransky scaling~\cite{Miransky:1984ef} (similar to the 
Berezinsky-Kosterlitz-Thouless scaling), read     
\begin{align} 
\alpha_g^{\rm cr} &=\frac{N_C}{4\pi} \cdot 
\frac{\pi}{3} \cdot \left(\frac{2N_C}{N_C^2-1} \right) = 
\frac{1}{6} \cdot \left(\frac{N_C^2}{N_C^2-1} \right) \simeq \frac{1}{6}
 \,, \notag \\ 
 m_F &\simeq 4 M_{\rm pl} \exp \left[- \frac{\pi}{\sqrt{ \frac{\alpha_g}{\alpha_g^{\rm cr}}} - 1} \right]
 \,, \label{Mira}
\end{align}
where the ultraviolet scale has been set to the 
Planck scale $M_{\rm pl}$.

As one reference, 
we may identify the size of the critical coupling as an 
AS fixed point $\alpha_g^*$ which has been predicted 
in the large $N_F$ gauge theory~\cite{Holdom:2010qs}~\footnote{
To be more precise, the validity of the large $N_F$ fixed point has been thoroughly 
questioned~\cite{Alanne:2019vuk}, 
where it is explained why the fixed point is an artefact of the
infinite $N_F$ limit, and incompatible with any finite set of higher-order corrections at finite $N_F$. 
This conclusion is also corroborated by advanced lattice simulations~\cite{Leino:2019qwk}, 
where no indications
for a UV fixed point of the above form have been found. 
At any rate, our main claim on the emergence of composite axion 
and the successful relaxation mechanism will not substantially be affected  
even if the gauge fixed point is unwarranted in the perturbative coupling regime. 
}: 
$\alpha_g^* = 3N_C/(2 N_F)$.  
In that case we would have 
\begin{align} 
N_F= 9 N_C \left( 1 - \frac{1}{N_C^2} \right) \simeq 9 N_C 
\,. \label{NC-NF}
\end{align}

\begin{figure}[t]
  \begin{center}
   \includegraphics[width=7cm]{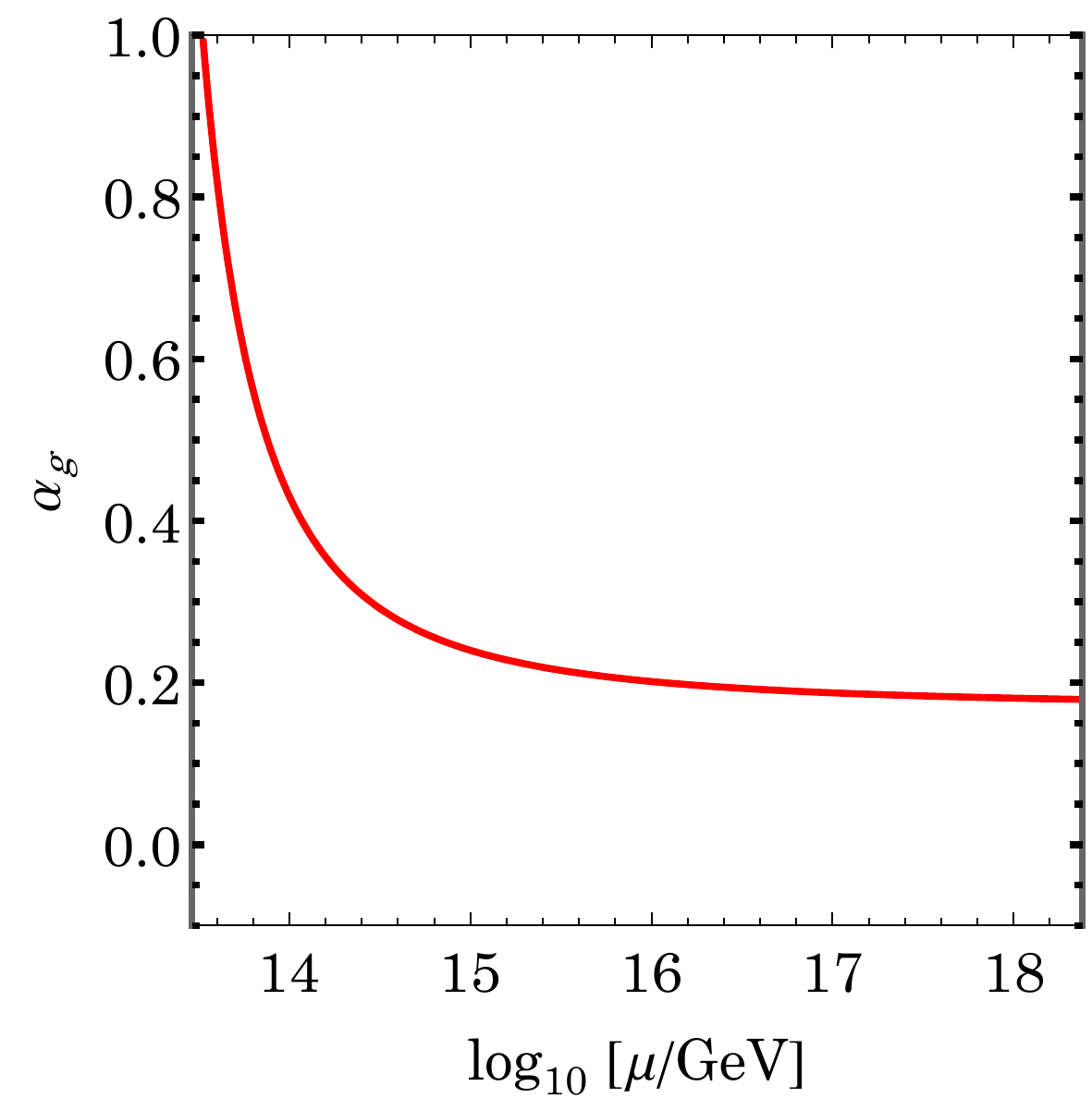}
  \end{center}   
\caption{
Renormalization evolution for axicolor ($\alpha_g$) coupling 
from the fermion dynamical mass scale $m_F$  
up to the Planck scale $M_{\rm pl} \simeq 2.4 \times 
10^{18}$ GeV 
in the strong coupling regime  for $\alpha_g$, in the anti-Veneziano limit 
($N_F, N_C \gg 1$ with $N_F/N_C \gg 1$), flowing to 
the AS fixed point predicted in the large $N_F$ gauge theory~\cite{Holdom:2010qs}. 
As a reference, 
the numbers of axicolors $(N_C)$ and axicolored fermions $N_F$ are taken as $(N_C, N_F)=(5, 45)$, inspired by Eq.(\ref{NC-NF}), 
and we have chosen   
$m_F = 2.9 \times 10^{13}$ GeV 
as in the text. 
At this scale the 
dimensional transmutation takes place following 
the nonperturbative beta function $\beta_g^{\rm NP}$ in Eq.(\ref{NP-beta}), 
or equivalently the Miransky scaling in Eq.(\ref{Mira}), 
and the scalar potential is expected to be stabilized all the way above 
this scale. 
}  
\label{RG-runs-nonpert}
\end{figure}

The Miransky scaling (the second relation in Eq.(\ref{Mira})) 
generates the nonperturtabutive scaling of $\alpha_g$ in the chiral 
broken phase (i.e. the nonperturbative scale anomaly) 
with respect to the cutoff $\Lambda=M_{\rm pl}$
~\cite{Leung:1985sn}: 
\begin{align} 
\beta_g^{\rm NP} &= - 2 \frac{\alpha_g^{\rm cr}}{\pi} \left( 
\frac{\alpha_g}{\alpha_g^{\rm cr}} - 1 \right)^{3/2}
\,, \notag \\ 
{\rm or} \qquad 
\frac{\alpha_g(\mu)}{\alpha_g^{\rm cr}} 
&= 1 + \frac{\pi^2}{\left[ \log \frac{\mu}{m_F} \right]^2} 
\,, \qquad  
{\rm for} \qquad 
\alpha_g > \alpha_g^{\rm cr}
\,. \label{NP-beta}
\end{align}
This nonperturbative running, which is actually ``walking'', 
governs the dynamics for $m_F < \mu < \Lambda=M_{\rm pl}$, 
with the characteristic large hierarchy $m_F \ll \Lambda=M_{\rm pl}$. 
Note that $\alpha_g^*$  indeed plays the role of the AS fixed point in Eq.(\ref{NP-beta}), 
and $m_F$ is the renormalization group invariant scale, like $\Lambda_{\rm QCD}$ 
in QCD, which signals the dimensional transmutation of the dynamical 
scale generation and is also the intrinsic scale characterizing the 
confining phase of the axicolor below $\mu = m_F$. 
In Fig.~\ref{RG-runs-nonpert} we draw the running coupling behavior in the chiral broken phase, in the walking regime $(m_F <\mu< \Lambda \equiv M_{\rm pl})$.


For 
$\mu < 2.9 \times 10^{13}$ GeV, 
we find $(\alpha_{\lambda_1} + \alpha_{\lambda_2})  <0$ which 
destabilizes the scalar potential at the two loop level~\footnote{
The scalar sector itself can have a fictitious AS fixed point 
in the large $N_F$ limit at the two loop level. 
However, it is
known on general grounds that scalar theories do not have fixed points at weak coupling~\cite{Bond:2018oco}. 
While fixed points might still exist at strong coupling, perturbation theory up to
four loops in the MS bar scheme is unable to clarify~\cite{Steudtner:2020tzo}. 
Specifically, the conjectured fixed points at two loop located
at $(\alpha_{\lambda_1},\alpha_{\lambda_2})=(0.333,-0.865)$ and $(\alpha_{\lambda_1},\alpha_{\lambda_2})=(0.333,0.642)$
disappear into the complex plane at three loop, while at four loop, they re-appear
at  $(\alpha_{\lambda_1},\alpha_{\lambda_2})=(0.199,-0.736)$ and $(\alpha_{\lambda_1},\alpha_{\lambda_2})=(0.199,1.774)$.
Results are not stable under the loop expansion, and couplings are not small enough
to be under perturbative control.  Therefore, the existence of a scalar fixed point 
cannot be trusted, hence the potential stability condition might not be reliable either. 
However, it will not essentially make our main claim on the emergence of composite 
axion and the successful relaxation altered, which holds in the nonperturbative 
strong axicolor dynamics, without referring to the details of the scalar dynamics.  
}. 
We may thus 
identify the intrinsic scale $m_F$ as the threshold scale: 
$ 
m_F \sim 2.9 \times 10^{13} \, {\rm GeV} 
$, 
which requires about 2\% deviation for $\alpha_g$ off $\alpha_g^*$ 
at $\mu = M_{\rm pl}$. 



\section{Identification of Composite Axion}

At $\mu = M_{\rm pl}$ 
the chiral $U(N_F)_L \times U(N_F)_R$ symmetry is already broken down to 
$U(N_F)_V$, by the dynamical mass generation in Eq.(\ref{Mira}) and 
developing nonzero chiral condensate 
$\langle - \bar{F}F \rangle$. 
However, axiclored fermions with the mass $m_F$ does not decouple to 
contribute to the renormalization-group running (walking) down until the intrinsic scale $m_F$, where the theory would be in the confining phase. 
At this scale, 
$N_F^2$ composite Nambu-Goldstone bosons are emergent. 
Among those Nambu-Goldstone bosons, 
the chiral singlet one 
($ \sim \bar{F} i \gamma_5 T^0 F$  
associated with the singlet 
generator, $T_0 = 1/\sqrt{2 N_F} \cdot {\bf 1}_{N_F \times N_F}$)   
gets mass of ${\cal O}(m_F)$, predominantly due to the axicolor instanton  via the $U(1)_{F_A}$ anomaly, like $\eta'$ in QCD.  
The colored Nambu-Goldstone bosons get masses of ${\cal O}(g_s/(4\pi) m_F)$ by 
the QCD gluon interaction with the gauge coupling $g_s$, analogously to 
the photon exchange contribution to the charged pions in QCD. 
The $n_F^2$ QCD color-singlets Nambu-Goldstone bosons   
are at this point massless.
Of those massless ones, we find the composite axion candidate, free from 
the axicolor $U(1)_{F_A}$ anomaly and coupled to the QCD topological operator $G_{\mu\nu} \tilde{G}^{\mu\nu}$:  
 \begin{align} 
a \sim \bar{F}  i \gamma_5 T_{a} F 
\label{Pia}
\end{align} 
with  the generator 
\begin{align} 
T_{a} &\equiv 
\sqrt{\frac{n_F}{6 N_F}}\left(
\begin{array}{cc} 
- 1_{3 \times 3} & 0 \\ 
 0 & \frac{3}{n_F} \cdot 1_{n_F \times n_F} 
\end{array}
\right) . 
\label{Ta}
\end{align}
This composite-axion identification is a generalization of 
the one in the axicolor theory with smaller $N_F$ ~\cite{Choi:1985cb}. 
The associated chiral symmetry protecting the masslessness for $a$ 
is identified as an anomalous PQ symmetry.

\section{Low-Energy Description below $m_F$}

Below $m_F$ 
all axi-hadrons would be decoupled, 
only leaving Nambu-Goldstone bosons (except $\eta'$ like one). 
Then we work on a chiral Lagrangian in the chiral broken phase 
based on $G/H = SU(N_F)_L \times SU(N_F)_R/SU(N_F)_V$ coupled to the 
chiral scalar $M$:   
\begin{align} 
{\cal L} 
 =& \frac{f^2}{4} {\rm tr}[D_\mu U^\dag D^\mu U] 
+ b \, f^2 \,  {\rm tr}[U^\dag {\cal  M} + {\cal M}^\dag U]
+ {\cal L}_{\rm WZW} 
\,, \label{Chiral-L}
\end{align}
with the suprion field ${\cal M}$ includes the chiral scalar $M$ as 
\begin{align} 
 {\cal M} = \left( 
 \begin{array}{cc} 
  0_{3 \times 3}& 0 \\ 
  0 & (M)_{n_F \times n_F}   
 \end{array}
 \right).  
 \label{calM}
\end{align}  
Here $U$ is the chiral field parametrized by the Nambu-Goldstone boson fields $\pi$ as 
$U  = e^{2 i \sum_{A=1}^{N_F^2-1} \frac{\pi^A T^A}{f}}$ with $T^A$ being $SU(N_F)$ generators 
and $f$ the axi-pion decay constant. 
This $U$  
transform under the chiral symmetry 
in the same way as $M$ does: $U \to g_L \cdot U \cdot g_R^\dag$ ($g_{L/R} \in 
SU(N_F)_{L/R}$).   
The covariant derivative acting on $U$ has been introduced by reflecting 
the partial-vectorlike gauging of the chiral symmetry 
as in Eq.(\ref{F}), which takes the form  
$ D_\mu U = \partial_\mu U - i [V_{\mu}, U]$, 
with the gauge field 
$ V_\mu  = 
 \left( 
 \begin{array}{cc} 
 g_s (G_\mu)_{3 \times 3} & 0 \\ 
 0 & 0_{n_F \times n_F}  
 \end{array}
 \right) 
$. 
The last part ${\cal L}_{\rm WZW}$, the (covariantized) Wess-Zumino-Witten term~\cite{Wess:1971yu,Witten:1983tw}, 
includes the QCD topological term, $G_{\mu\nu}^{\bf a} \tilde{G}^{{\bf a} \mu\nu}$ (with ${\bf a} = 1, \cdots , 8$),  
coupled to the composite axion $a$ in Eq.(\ref{Pia}): 
\begin{align} 
 {\cal L}_{\rm WZW} 
 &\ni 
 - \frac{N_{C}}{4 \pi^2 f} \epsilon^{\mu\nu\rho\sigma} {\rm tr} 
  [\partial_\mu V_\nu \partial_\rho V_\sigma \pi] 
\notag \\ 
& \ni 
\frac{N_{C} g_s^2}{32 \pi^2 }  \frac{a}{f_a} G^{\bf a}_{\mu\nu} \tilde{G}^{{\bf a} \mu\nu}
\,, \label{aGGtilde}
\end{align}
with the axion decay constant 
\begin{align} 
f_a = \sqrt{\frac{6 N_F}{n_F}} f
.\label{fa}
\end{align}

The coupling parameter $b$ in Eq.(\ref{Chiral-L}) can be fixed as follows: 
replace the $\psi$- fermion bilinear $\bar{\psi}_{R i} \psi_{L j}$ in Eq.(\ref{Ly}) 
with a partial chiral field $u_{ji }$ as $\bar{\psi}_{R i} \psi_{L j} \approx 
\langle \bar{\psi} \psi \rangle \cdot u_{ji} = 
\langle \bar{F} F \rangle \cdot u_{ji}$, 
where 
we have introduced the $SU(N_F)$ vectorial chiral condensate per flavor,  
$\langle \bar{\psi} \psi \rangle = \langle \bar{F} F \rangle$, 
which is respected in the axicolor-vectolike gauge dynamics via 
the Vafa-Witten's theorem~\cite{Vafa:1983tf}.  
Taking $u=1$ in the $b$ term in Eq.(\ref{Chiral-L}), and comparing them, 
we then find 
\begin{align} 
 b = \frac{ y \langle - \bar{F}F \rangle}{f^2}
\,. \label{b}
\end{align}

Including the quartic-potential terms in Eq.(\ref{VM}), 
 the potential of $M$ reads  
\begin{align} 
V_M(\mu=m_F) 
= &
-  b \, f^2 {\rm tr}[M + M^\dag] 
\notag \\ 
& +  
\lambda_1 {\rm tr}[(M^\dag M)^2] + \lambda_2 ({\rm tr}[M^\dag M])^2
\,. \label{VM-mF}
\end{align}
Thus the chiral condensate ($b$ in Eq.(\ref{b})) generates 
the tadpole term for the chiral scalar $M$. 
Note that though the chiral condensate is generated already at 
$\mu=M_{\rm pl}$, the theory keeps (almost) scale-invariant down til $\mu \sim m_F$ 
due to the walking nature, as seen from the $\alpha_g$ running in Fig.~\ref{RG-runs-nonpert}. 
Thereby, the scalar potential in Eq.(\ref{VM-mF}) has been defined 
at $\mu=m_F$, hence the tadpole term coefficient $(b f^2)$ should 
be read as $(b f^2)|_{\mu = m_F} = y(m_F) \langle - \bar{F}F \rangle|_{\mu = m_F}$, 
via Eq.(\ref{b}).

Decomposing $M$ into real and imaginary parts as 
$
M = S + i P$, 
the potential 
looks like 
\begin{align} 
V_M(m_F) = 
-  
\sqrt{2 n_F} b f^2 \cdot S_0 +  
\frac{1}{4}\left( \frac{\lambda_1}{n_F} + \lambda_2 \right) \cdot S_0^4, 
\end{align}
where only the $SU(n_F)$-singlet scalar part ($S_0 \cdot 1/\sqrt{2 n_F} \cdot 1_{n_F \times n_F}$) has survived, 
because others do not get the tadpole. 
We then see that the chiral scalar develops nonzero vacuum expectation value (VEV), 
\begin{align} 
\langle M \rangle & 
= \langle S_0\rangle \cdot \frac{1}{\sqrt{2 n_F}} \cdot 1_{n_F \times n_F}
= \left( \frac{b f^2}{2 (\lambda_1 + n_F \lambda_2)} \right)^{1/3} \cdot 1_{n_F \times n_F} 
\notag \\ 
& = 
\left( \frac{ n_F b f^2}{2 (4 \pi)^2 (\alpha_{\lambda_1} + \alpha_{\lambda_2})} \right)^{1/3} \cdot 1_{n_F \times n_F} 
\equiv v_M \cdot 1_{n_F \times n_F}, 
\label{MVEV}
\end{align} 
where we have replaced $\lambda_1$ and $\lambda_2$ with their fine structure constants, in Eq.(\ref{alphas}). 

Note that this VEV does not generate any extra Nambu-Goldstone bosons: 
turning off the Yuakwa coupling $y$ in Eq.(\ref{Ly}) (as well as
 QCD charges), the theory possesses an enhanced global chiral $[U(N_F)_L \times 
 U(N_F)_R] \times [U(n_F)_L \times 
 U(n_F)_R]$ symmetry. In this sense, nonzero $y$ explicitly breaks 
 the enhanced chiral symmetry down to single $[U(n_F)_L \times 
 U(n_F)_R]$, which is spontaneously 
 broken by the $F$-fermion chiral condensate in the axicolor sector. 
 As seen above, this chiral condensate induces the tadpole ${\rm tr}[S_0\cdot 1_{n_F \times n_F}]$, which only respects the vectorial $U(n_F)_{F_V}$ symmetry, 
 and it indeed explicitly breaks the original $U(n_F)$ chiral symmetry in the $M$ 
 sector. Hence pseudoscalars in $M$ fully get massive by the $y$-induced tadpole, and acquire the same mass as the $(n_F-1)$-plet scalars, $m_P^2 = \frac{2 (4\pi)^2}{n_F} (2 \alpha_{\lambda_1} + \alpha_{\lambda_2}) \cdot v_M^2$. Some of them also gets the mass from the axicolor and/or QCD 
 axial anomalies. 
 Thus no extra massless Nambu-Goldstone bosons emerge. 

The chiral scalar VEV $\langle M  \rangle$ backreacts on the $F$-fermion mass as the chiral explicit breaking effect, which in part includes 
the PQ symmetry breaking. 
We in fact see that 
the chiral Lagrangian in Eq.(\ref{Chiral-L}) gives the Nambu-Goldstone boson potential terms:  
\begin{align} 
 V_y &= 
- b f^2  {\rm tr}[ U^\dag \langle  {\cal M} \rangle + {\rm h.c.}] 
\notag \\ 
&=
- b f^2  {\rm tr} \left[ U^\dag 
\left( 
\begin{array}{cc}
0_{3 \times 3} & 0 \\ 
0 & (M)_{n_F \times n_F}  
\end{array} 
\right)
+ {\rm h.c.} 
\right]
\notag \\ 
& =
- b f^2 v_M  {\rm tr} \left[ 
e^{ - 2i \left( \frac{\pi_0}{\sqrt{2 N_F} f} + 3 \sqrt{\frac{n_F}{6N_F}} \frac{\pi_a}{n_F f} \right)} \cdot 1_{n_F \times n_F}  \cdot
e^{2i \frac{P_0}{\sqrt{2 n_F} v_M}} \cdot 
1_{n_F \times n_F}    
+ {\rm h.c.} 
\right] + {\rm others}
\notag\\ 
& =
- 2 n_F b f^2 v_M   
\cos \left( 
2  \frac{\pi_0}{\sqrt{2 N_F} f} + 6 \frac{\pi_a}{n_F f_a}  
- \frac{2 P_0}{\sqrt{2 n_F} v_M} \right) 
  + {\rm others}
\,,\label{VU}
\end{align} 
where we have parametrized $M$ as nonlinear realization, $M = v_M \cdot e^{2i P/v_M}$ with 
$v_M $ being chosen to be positive, and  
terms denoted by others do not contribute to the axicolor and QCD anomalies, 
hence can be taken to be zero in evaluating minimization of the potential. 
Then we find the potential term for $\pi_0 \equiv \eta_{\rm AC}$ and $\pi_a \equiv a$ 
(normalized as $V_y(\eta_{\rm AC}=0, a=0, P_0 =0)=0$), 
\begin{align} 
 V_y(\eta_{\rm AC}, a, P_0) 
=&  2 n_F b f^2 v_M 
\Bigg[ 
1 - \cos \left( 2  \frac{\eta_{\rm AC}}{\sqrt{2 N_F} f} + \frac{6}{n_F}  \frac{a}{f_a}  
- \frac{2 P_0}{\sqrt{2 n_F} v_M}
\right) 
\Bigg] 
\,.   \label{Va}
\end{align}
As shown in Appendix A, no phase parameter arises from the axicolor sector 
in this potential, due to masslessness of QCD-colored fermions. 
Note that the potential terms in Eq.(\ref{Va}) necessarily come with the suppression factor of $1/n_F$ or $1/N_F$. 
Thus, the $y$-Yukawa interaction breaking the PQ symmetry generates the potential terms for  
the axion as well as the potential-mixing partners $\eta_{\rm AC}$ and $P_0$,  
as the subleading-order effect in the anti-Veneziano limit including the 
large $N_F$ expansion.

 \section{Relaxation and Axion Mass}

In addition to the potential terms in Eq.(\ref{Va}), 
the quantum anomalies give rise to the corrections: 
from the axicolor sector (with the axicolor gauge field ${\cal G}$ with the gauge coupling $g_{\rm AC}$), we have 
\begin{align} 
{\cal L}_{{\cal G} \tilde{\cal G}} 
= \frac{g_{\rm AC}^2}{32\pi^2} 
\left( \frac{\eta_{\rm AC}}{\sqrt{2 N_F} f} + \frac{P_0}{\sqrt{2 n_f} v_M}   \right) 
{\cal G}_{\mu\nu}^A \tilde{\cal G}^{\mu\nu A} 
\,. 
\end{align}
The axicolor instanton therefore yields the potential,  
\begin{align} 
V_{{\cal G} \tilde{\cal G}}^{\rm instanton}  
\sim m_F ^4  
\left[ 1 - \cos \left( \frac{\eta_{\rm AC}}{\sqrt{2 N_F} f} + \frac{P_0}{\sqrt{2 n_f} v_M}   \right) \right] 
\,. 
\end{align}
Again, the potential terms arise as the subleading effect in the large $N_F$ expansion, suppressed 
by $1/n_F$ or $1/N_F$.

Besides, the QCD sector develops the anomaly along with the theta parameter $\theta$, 
\begin{align}
{\cal L}_{G\tilde{G}}
=
\frac{g_s^2}{32 \pi^2 }  
\left( 
N_{C}  \frac{a}{f_a} 
+ 
N_C \frac{\eta_{\rm AC}}{\sqrt{2 N_F} f}
+ \theta 
\right)
G^{\bf a}_{\mu\nu} \tilde{G}^{{\bf a} \mu\nu}
\,,  
\end{align}
with which the QCD instanton generates the potential terms, 
\begin{align}
V_{G\tilde{G}}^{\rm instanton}
\sim
\left( \Lambda_{\rm QCD}^{\rm flavor-singlet} \right)^4  
\left[1 - \cos \left( 
N_{C}  \frac{a}{f_a} 
+ 
N_C \frac{\eta_{\rm AC}}{\sqrt{2 N_F} f}
+ \theta 
\right)
\right] 
\,. 
\end{align}
The overall scale $\Lambda_{\rm QCD}^{\rm flavor-singlet}$ denotes 
a flavor-singlet part of the QCD instanton contribution, such as 
$\sim m_u m_d \Lambda_{\rm QCD}^2$ for the lightest two quark flavors, arising from 
the flavor-singlet nature of the QCD vacuum (See, e.g., a review~\cite{Kim:2008hd})~\footnote{
The QCD $\eta'$ would mix there, as discussed in~\cite{Kim:2008hd}, 
but yield the mass from another instanton-induced potential energy $\sim \Lambda^4_{\rm QCD}$, irrespective 
to the flavor-singlet condition. Incorporation of the QCD $\eta'$ into 
the minimization of the full potential would not substantially modify Eqs.(\ref{relaxation-condi}) 
and (\ref{relax}). 
}. 
The $\eta_{\rm AC}$ terms are suppressed by factors of $1/N_F$, while 
the axion terms survive even in the leading order of the large $N_F$ expansion. 
This is due to the fact that the axion coupling to $G_{\mu\nu} \tilde{G}^{\mu\nu}$ 
arises from the QCD-charged $\psi^\alpha$ fermion loop which does not 
come with the $(1/n_F)$ factor (See Eqs.(\ref{aGGtilde}) with Eq.(\ref{Ta})). 
Note that even in terms of the anti-Veneziano limit ($N_C, N_F \gg 1, N_F/N_C \ll 1$), 
only the axion term together with the QCD $\theta$ parameter survives.  

Combining those with Eq.(\ref{Va}), 
we find the full potential to be minimized by the following condition:   
\begin{align} 
 \left( \frac{\eta_{\rm AC}}{\sqrt{2 N_F} f} + \frac{P_0}{\sqrt{2 n_f} v_M}   \right) 
& = 
 \left( 
N_{C}  \frac{a}{f_a} 
+ 
 N_C \frac{\eta_{\rm AC}}{\sqrt{2 N_F} f}
+ \theta 
\right)
\notag\\ 
& = 
 \left( 2  \frac{\eta_{\rm AC}}{\sqrt{2 N_F} f} + 6  \frac{a}{n_F f_a}  
- \frac{2 P_0}{\sqrt{2 n_F} v_M}
 \right) 
=0 
\,. \label{relaxation-condi}
\end{align}
In the anti-Veneziano limit, i.e., $N_F \gg 1, N_C \gg 1$ and $N_F/N_C \gg 1$, only the $a$ - $G$ - $\tilde{G}$ coupling is left so that 
the minimization condition is simplified to be  
\begin{align} 
  \frac{a}{f_a} \stackrel{N_F \gg 1}{\approx} - \frac{\theta}{N_C} 
\,. \label{QCD-axion}
\end{align}
This is precisely the limit for the (composite) QCD axion feeding the mass only through 
the QCD instanton effect, which is essentially identical to the one in the original axicolor scenario~\cite{Kim:1984pt,Choi:1985cb} without a $y$-Yukawa coupling as in Eq.(\ref{Ly}). 
However, this case is too extreme to be phenomenologically viable, because all the $y$-Yukawa term contributions to the potential are suppressed by $1/N_F$, leaving a number of exactly massless Nambu-Goldstone bosons.

Including the subleading terms in the large $N_F$ expansion, 
from Eq.(\ref{relaxation-condi}) the relaxation of the theta parameter reads 
\begin{align} 
 \frac{a}{f_a} &= - \frac{\theta}{N_C \left( 1 - \frac{3}{2 n_F} \right)}
\,, \notag\\ 
 \frac{\eta_{\rm AC}}{\sqrt{2 N_F} f} & = - \frac{3}{2} \frac{a}{f_a} 
\,, \notag\\ 
 \frac{P_0}{\sqrt{2 n_F} v_M} & = - \frac{\eta_{\rm AC}}{\sqrt{2 N_F} f} 
\,. \label{relax}
\end{align}
This relaxation works including heavier background fields other than the composite 
axion, which 
is called ultraviolet solutions as recently discussed in the literature~\cite{Draper:2016fsr,Agrawal:2017ksf,Gaillard:2018xgk,Gherghetta:2020ofz}.  We will come back to this point again later. 
Thus, the viable composite axion with the successful relaxation mechanism arises 
as the subleading effect in the large $N_F$ expansion, i.e., the subleading correction 
to the anti-Veneziano limit.

At $\mu = M_{\rm pl}$, 
the chiral symmetry of the axicolored fermions  
is only broken in part by the $y$-Yuakwa interaction with 
the chiral scalar $M$. 
No higher dimensional interactions, including gravitational chiral breaking,  
are allowed by the quantum scale invariance at this Planck scale, 
robustly protected 
by the AS fixed point. 
The partially gauged perturbative QCD interactions cannot 
induce 
any chiral breaking at any loop level, because it is protected 
by the chiral symmetry.

The $y$-Yukawa interaction will not generate any 
chiral breaking, as long as the Yukawa coupling is perturbative enough, 
as we have assumed 
(e.g. $10^{-90} \lesssim \alpha_y(m_F) \lesssim 10^{-6}$, for a later reference).
Thus the $y$-Yukawa interaction will be frozen at $m_F$ because 
of decoupling of axicolor fermions,  

Even over the Planck scale, 
the present scalegenesis can be protected by an asymptotic safety 
quantum gravity with the nontrivial ultraviolet fixed point~\cite{Gies:2003dp,Shaposhnikov:2008xi,Gies:2009sv,Braun:2010tt,Bazzocchi:2011vr,Wetterich:2011aa,Antipin:2013pya,Gies:2013pma,Tavares:2013dga,Abel:2013mya,Bond:2016dvk,Pelaggi:2017abg,Bond:2017wut,Barducci:2018ysr,Eichhorn:2018yfc,Abel:2018fls}. 
Thereby, any quantum gravity contribution to the chiral symmetry breaking, arising as higher dimensional operators suppressed by the Planck scale, 
could be absent as well (See also footnote~\ref{foot2}). 
Thus the theory could  keep (nearly) scale invariant all the way in running (``walking") 
down until 
the axicolor undergoes the dimensional transmutation at $m_F$, 
due to the walking nature in the chiral broken phase 
along with the Miransky scaling (Eq.(\ref{Mira}) or Eq.(\ref{NP-beta})). 
This implies that the theory could have no quality problem~\cite{Holman:1992us,Kamionkowski:1992mf,Barr:1992qq,Ghigna:1992iv}.

Though being frozen, 
the remnant of $y$-Yuakwa interaction will be crucial only after the scale generation at $m_F$, beause it will trigger 
the chiral scalar VEV,  and feed back to the axicolored fermion masses. 
This develops the composite axion potential as in Eq.(\ref{Va}). 
Actually, this phenomenon arises as a subleading effect in terms of 
the present asymptotic safety (with the large $N_F$ and $N_C$),
 as noted above. 
So, the axion (and other pseudo Nambu-Goldstone bosons) will be well under control by 
perturbation around the anti-Veneziano limit, as well as the standard chiral 
perturbation theory as in the case of QCD pion and/or axion. 
Thus the $y-$ Yukawa induced axion potential form in Eq.(\ref{Va}) will be intact, as long as 
the chiral and the AS protection works well. 

Moreover, of interest is to note that 
the axicolor-theta term is unphysical in the present setup, 
because QCD-colored $\psi^\alpha$ fermions do not couple to 
the chiral scalar $M$, and do not get the mass directly by the $y$ coupling. 
Note that this imbalance between double chiral symmetries in 
the fermion and scalar sectors ($N_F$ and $n_F=N_F-3$) has nothing  
essentially to do with realization of asymptotic safety (large $N_F$ walking feature). 
Thus  
the axicolor-theta term can completely be rotated away by 
the residual chiral symmetry. 
Hence the potential form of Eq.(\ref{Va}) is unambiguous under 
the axial rotation in the axicolor sector, as it stands. 
More details are provided in Appendix A. 

Thus the composite axion relaxation mechanism for the QCD strong 
CP phase perfectly works, 
supported by the asymptotic safety (large $N_F$ walking), 
which is already fulfilled at the infrared dynamical scale $m_F$, much higher 
than the QCD scale, where the renormalization evolution of the $\theta$ 
arising from the standard model alone, 
by scaling down to the QCD scale will be negligibly small. 
Thus, the strong CP problem is solved at such an ultraviolet scale by 
a hybrid relaxation mechanism involving other heavier psuedoscalars, 
as in Eq.(\ref{relax}). 
Similar ultraviolet solutions in a context of composite axions 
have been discussed in the literature~\cite{Draper:2016fsr,Agrawal:2017ksf,Gaillard:2018xgk,Gherghetta:2020ofz}.

By integrating out heavier pseudoscalars ($\eta_{\rm AC}$ and $P_0$) with mass predominantly fed by the axicolor instanton effect, 
the total axion mass (squared) reads  
\begin{align} 
 m_a^2 \simeq 
 \frac{  (\Lambda_{\rm QCD}^{\rm flavor-singlet})^4 + \frac{72}{n_F} \cdot m_{F0} \langle - \bar{F}F \rangle 
  }{f_a^2} 
  \equiv  (m_a)_{\rm QCD}^2 + (m_a)^2_{\rm axicolor} 
\,. \label{ma}
\end{align}
where $m_{F0} \equiv y v_M$. 
Now we discuss the generic property of $(m_a)_{\rm axicolor}$ 
in the anti-Veneziano asymptotic safety. 
To this end, note first that the axicolor-sector vacuum-energy part,  
$m_{F0} \langle - \bar{F}F \rangle$ in Eq.(\ref{ma}) is renormalization group invariant. 
We set the reference renormalization scale $\mu$ to $m_F$.  
Then we can generically evaluate 
the chiral condensate per flavor (renormalized at $\mu=m_F$) and 
the axi-pion decay constant $f$ (and $f_a$ through 
Eq.(\ref{fa})), 
as a function of $m_F$: 
\begin{align} 
 \langle - \bar{F}F \rangle_{\mu =m_F} & = 
 \frac{N_C}{4 \pi^2} \kappa_C m_F^3 
 \,, \notag \\ 
 f^2 & = \frac{N_C}{4 \pi^2} \kappa_F m_F^2 
 \,,  
 \label{fpi-cond}
\end{align}  
up to constants $\kappa_F$ and $\kappa_C$ which can be fixed by solving 
the walking dynamics in the chiral broken phase. 
When we use the ladder Schwinger-Dyson equation along with 
the Pagels-Storkar formula~\cite{Pagels:1979hd}, 
these $\kappa_F$ and $\kappa_C$ can be estimated to be~\cite{Miransky:1989qc,Matsuzaki:2015sya} 
\begin{align} 
\kappa_F & \simeq 2 \xi^2 \simeq  2.4
\,, \notag \\ 
 \kappa_C &\simeq\frac{32 \xi^2}{\pi^2} \simeq 3.9   
\,, \label{LAD-PS}
\end{align}
where $\xi \simeq 1.1$. 
Then a reference value of the axion decay constant $f_a$ in Eq.(\ref{fa}) is estimated 
by taking account the aforementioned stability bound for the scalar potential 
in the walking regime 
$(m_F \sim 2.9 \times 10^{13} \, {\rm GeV})$ 
as 
$ f_a \sim 10^{14}\,{\rm GeV} 
$, 
for $N_F=9N_C=45$.

Using Eq.(\ref{fpi-cond}) together with Eq.(\ref{MVEV}) and replacing $y$ with its fine structure constant $\alpha_y$ in Eq.(\ref{alphas}), 
we express the axicolor-induced axion mass in Eq.(\ref{ma}) by 
making the $N_C$ and $N_F$ dependence manifest: 
\begin{align} 
 \left( \frac{(m_a)_{\rm axicolor}}{m_F} \right)^2 
 &= \frac{12}{N_F} \left(  \frac{\kappa_C}{\kappa_F}  \right) \cdot 
 \left( \frac{  [m_{F0}]_{\mu=m_F} }{m_F} \right) 
 \notag \\ 
 & 
 = \frac{12}{N_F} \left(  \frac{\kappa_C}{\kappa_F}  \right) \cdot 
 \left[  \frac{2 n_F \alpha_y^2}{N_C (\alpha_{\lambda_1} + \alpha_{\lambda_2})}   \right]^{1/3}_{\mu=m_F}
\,.\label{ma-mF}
\end{align}
Note that $m_F$ does not scale with $N_C$ or $N_F$. 
Thus $(m_a)_{\rm axicolor}$
 is potentially a subleading term of ${\cal O}(1/\sqrt{N_F})$
 and becomes vanishingly small in the anti-Veneziano limit, 
 \begin{align}  
& (m_a)_{\rm axicolor}\to 0\,, \qquad {\rm i.e.}\,,\qquad  m_a \to (m_a)_{\rm QCD} 
\,, 
\notag \\ 
{\rm as} \qquad 
& 
 N_F, N_C \to \infty\,,  
\qquad {\rm with} \quad N_F/N_C 
\gg 1 
\,,    
\label{ma-scaling}
\end{align}
in accordance with the same relaxation form as the QCD axion's, in Eq.(\ref{QCD-axion}). 
This feature is contrast to the $\eta'$ like mass in the anti-Veneziano limit, 
which goes like $\gg m_F$, as was discussed in the literature~\cite{Matsuzaki:2015sya}.

With Eq.(\ref{ma-mF}) for finite $N_F$ and $N_C$ (e.g. $N_F=9 N_C=45$), the reference values may be in a range 
involving 
$m_a \sim (m_a)_{\rm QCD} \sim 10^{-7}$ eV for $\alpha_y(m_F) \sim 10^{-90}$, 
and $m_a \sim 10^{12}$ GeV for $\alpha_y(m_F) \sim 10^{-6}$.  
The former is the QCD axion limit with $f_a \sim 10^{14}$ GeV, 
while the latter is just a conservative upper bound for 
the current renormalization group analysis neglecting $\alpha_y$ 
compared to the perturbative scalar quartic couplings 
which we take $(\alpha_{\lambda_1} + \alpha_{\lambda_2})|_{\mu =m_F} \simeq 1.1 \times 10^{-5}$ from the two-loop level analysis~\footnote{
The two-loop perturbative running with this reference value of $(\alpha_{\lambda_1} + \alpha_{\lambda_2})|_{\mu =m_F}$ grows up to higher scales, until  
the fixed point $(\alpha_{\lambda_1},\alpha_{\lambda_2})=(0.333,0.642)$, hence no destabilization 
of the scalar potential is all the way. 
}. 
In the QCD axion limit, the composite axion with $f_a \sim 10^{14}$ GeV 
is currently in an unexplored window~\cite{Zyla:2020zbs}.

\section{Conclusion and Discussions}

In this paper, we proposed a new class of a composite axion 
providing a solution to the strong CP problem, which simultaneously 
resolves the hierarchy and triviality problems potentially 
existing in the PQ symmetry breaking scenario.  
The composite axion arises in a dynamical scalegenesis, and shows up in 
a subleading domain 
of an ``anti-Veneziano'' limit of 
a confining gauge theory 
($N_F,N_C \to \infty$, with $N_F/N_C  \gg 1$)   
having the an AS fixed point. 
The confining gauge theory, what we call axicolor, 
couples to $N_F$ fermions, three of which are gauged by the QCD charge and the rest $n_F(=N_F -3 \gg 1 )$ fermions 
couple to $n_F^2$ scalars to ensure the chiral symmetry for $n_F$ fermions 
by forming the Yukawa coupling in a scale invariant way.

We addressed that in the strong-gauge coupling regime, 
the large $N_F$ axicolor theory is 
governed by much like a strong QED with the constant gauge coupling identified as the AS coupling, that is a walking dynamics.
The full chiral symmetry for $N_F$ axicolored fermions, 
as well as the scale symmetry,  
are spontaneously broken by the dynamical fermion mass generation following 
the Miransky scaling (Eq.(\ref{Mira})). 
This chiral breaking generates an axion-like pseudoscalar composite 
and the Yukawa coupling with the developed scalar VEV serves as  
the axion mass, as well as the other pseudo Nambu-Goldstone boson's.

It was shown that 
in the anti-Veneziano limit, the composite axion becomes the QCD axion, 
and the relaxation of the QCD theta parameter works in the same way
as in the QCD axion's (Eqs.(\ref{QCD-axion}) and (\ref{ma-scaling})). 
The subleading nature of the anti-Veneziano limit makes 
the scenario more phenomonologically viable, and yields  
a heavy composite axion (Eq.(\ref{ma-mF})).

The present relaxation of the strong CP phase 
involves the so-called ultraviolet solution, along with 
a couple of heavy (composite) pseudoscalars, as well as the composite axion 
(Eq.(\ref{relax})).  
This is now interpreted as the subleading 
phenomenon of the anti-Veneziano-AS axicolor theory, 
and is robustly established (near) scale invariance during 
the walking regime (Fig.~\ref{RG-runs-nonpert}).

In closing, we give several comments on the directions 
along the presently proposed scenario, which are to be deserved to the future works: 
\begin{itemize}

\item The present gauge-Yukawa-scalar 
theory can be projected up to the gauged Nambu-Jona Lasinio theory, 
as has been noted in~\cite{Rantaharju:2017eej,Krog:2015bca}. 
In that case, the chiral scalar $M$ would be regarded as composites formed 
as the axicolored fermion bound states generated by the strong 
gauge and four-fermion interactions~\cite{Krog:2015bca}. 
This gauged Nambu-Jona Lasinio theory can then be 
almost scale-invariant until reaching the ultraviolet fixed point (along 
the ``fixed line" in the gauge and four-fermion coupling space~\cite{Kondo:1988qd,Appelquist:1988fm,Kondo:1991yk,Kondo:1992sq,Kondo:1993ty,Kondo:1993jq,Harada:1994wy,Kubota:1999jf}).  
It would be worth revisiting  
the gauged Nambu-Jona Lasinio theory in light of  
the emergence of the composite axion. 
In particular, in that case all the factors relevant to 
the axion mass formula in Eq.(\ref{ma-mF}) 
can be computed, so it would be possible to estimate 
more definite referenced value of $m_a/m_F$.

\item 
We have presently focused only on the asymptotically safe dynamical scalegenesis for 
the PQ sector. Extension of this scenario to the electroweak sector, i.e., 
the scalegenesis of the electroweak scale, might be straightforward simply by 
introducing the (currently suppressed) Higgs-portal coupling, like 
$(H^\dagger H)({\rm tr}[M^\dagger M])$. 
In that case, however, the size of such a portal coupling needs to be 
taken to be so tiny, which would revive the gauge-hierarchy problem. 
One possible way out is to address the dynamical origin of the chiral scalar $M$, 
as commented in the first bullet above, that is embedding the present 
liner sigma model description for the chiral scalar $M$ by the gauged Nambu-Jona Lasinio 
theory. 
The Higgs portal coupling is then absent at the leading order of the large $n_F$ expansion, 
because the QCD-singlet axicolor fermions do not couple to $H^\dag H $ at this level. 
Including the SM contribution as the subleading order of the large $n_F$, 
the portal coupling could be generated as small violation of  
the scale invariance. 
It would be highly suppressed and generated below 
the axicolor scale $m_F$, by involving 
the SM quark and gluon loops at higher loop levels of perturbation in the SM with 
the chiral scalar $M$. 
Taking into account the renormalization evolution of 
the induced portal coupling, scaling down to the electroweak scale from 
the axicolor scale $m_F$, might realize a desired size of the negative-squared 
Higgs mass parameter necessary to generate the Higgs VEV at the electroweak scale.  
This issue is noteworthy, to be pursued in details elsewhere.

\item 
The domain wall problem triggered by the axion 
is avoidable if an inflation takes 
place after the dynamical chiral symmetry breaking 
triggered by the axicolor (and four-fermion) dynamics. 
An inflaton can be played by a composite dilaton, 
arising as a pseudo Nambu-Goldstone boson associated with 
the spontaneous scale symmetry breaking due to the walking (i.e. nearly scale invariant) dynamics of the axicolor, hence  
is expected to be light enough to present in the low-energy 
dynamics together with the axi-pions and composite axion. 
As has been shown in~\cite{Ishida:2019wkd}, 
the walking dynamics can naturally achieve 
a small field inflation with the light walking dilaton as the inflaton. 
Or, as was discussed in~\cite{Inagaki:2015eza,Inagaki:2016vkf,Inagaki:2017ymx}, 
in the framework of the gauged Nambu-Jona Lasinio theory 
which could be dual to the gauge-Yukawa-scalar theory, 
allowing a nonminimal coupling to gravity 
can achieve a large field inflation, which was 
shown to favor somewhat a large $N_F$. 
For both cases, it would be interesting to examine 
how the composite axion can contribute to ending the 
inflation, i.e., so-called the waterfall scenario~\cite{Linde:1993cn,Gong:2021zem}.

\item 
Axicolor hadrons as well as the chiral scalars 
are heavy enough to be decoupled at low energy, except the lighter 
composite pseudo Nambu-Goldstone bosons (including a 
possibly light composite dilaton).  
Other than the composite axion,  
those pseudoscalars get the mass 
on the same order as $m_a$, arising at the subleading 
order of the large $N_F$ limit. 
They would be dark matters, and can be thermally produced 
by axi-hadronic interactions, 
such as the axi-baryon annihilations and/or $3-2$ processes 
like dark pions (strongly interacting massive particle, SIMP)~\cite{Hochberg:2014dra,Hochberg:2014kqa,Hochberg:2015vrg}.  
More on this dark matter physics, in association with the composite 
axion, would be worth to pursue.

\item 
For the present proposal to be more realistic,  
it needs more rigorous discussion on certain existence of asymptotic safe confining gauge-Yukawa-scalar theory 
in the nonperturbative regime, by using the functional 
renormalization group~\cite{Polchinski:1983gv,Wetterich:1992yh,Morris:1993qb,Litim:2001up}
 and lattice simulations as the pioneering work in~\cite{Leino:2019qwk}.

\end{itemize}

\section*{Acknowledgements}

S.M. work was supported in part by the National Science Foundation of China (NSFC) under Grant No.11747308, 11975108, 12047569 and the Seeds Funding of Jilin University. 
The work of H.I. was partially supported by JSPS KAKENHI Grant Number 18H03708.

\appendix 

\section{Massless Fermion Solution for Axicolor CP Violation}

This appendix shows that 
the axicolor sector is CP violating-phase free due to the 
flavor-singlet nature of the axicolor.

We focus on the anomalous sector:  
\begin{align} 
 {\cal L}_{U(1)_{F_A}, U(1)_{\rm PQ}}^{\rm anomalous} 
 &= 
  - \sum_{i=1}^{N_F}|m_i| \left[ e^{\i \theta_y} \bar{F}_{Li} F_{Ri} + {\rm h.c. } 
  \right] 
  + \frac{\theta_{\rm axicolor}}{32 \pi^2} {\cal G}_{\mu\nu} \tilde{\cal G}^{\mu\nu} 
  + \frac{\theta_{\rm QCD}}{32 \pi^2} {G}_{\mu\nu} \tilde{G}^{\mu\nu} 
\,.
\end{align}
where $\theta_{\rm QCD} = \theta - {\rm argdet}[m_q]$ with the QCD theta parameter $\theta$ 
and complex quark mass matrix $m_q$,  $m_1=m_2=m_3=0$ and $m_{4}=\cdots = m_{N_F}=m$.  
First of all, we note that the phase of the $y$-Yukawa coupling 
can be removed by field redefinition of $F_L$ and $F_R$, like 
$F_L \to e^{i\theta_y} F_L$ and $F_R \to e^{i \theta_y} F_R$, which keeps  
the other terms in the theory being invariant. 
Therefore we start with the phase-less Yukawa term: 
\begin{align} 
 {\cal L}_{U(1)_{F_A}, U(1)_{\rm PQ}}^{\rm anomalous} 
 &= 
  - \sum_{i=1}^{N_F}|m_i| \left[ \bar{F}_{Li} F_{Ri} + {\rm h.c. } 
  \right] 
  + \frac{\theta_{\rm axicolor}}{32 \pi^2} {\cal G}_{\mu\nu} \tilde{\cal G}^{\mu\nu} 
  + \frac{\theta_{\rm QCD}}{32 \pi^2} {G}_{\mu\nu} \tilde{G}^{\mu\nu} 
\,.
\end{align}

Under the $U(1)_{F_A}$ and $U(1)_{\rm PQ}$ rotations, the $F$-axicolored fermion fields 
transform as 
\begin{align} 
 F_{i} &\to [e^{- i\gamma_5 \alpha_{0} T_0/2}]_{ij} \cdot F_j 
 \qquad {\rm with} \qquad T_0 = \frac{1}{\sqrt{2 N_F}} {\rm diag}\{1,\cdots,1  \}
\,, \notag\\ 
 F_{i} &\to [e^{- i\gamma_5 \alpha_{a} T_a/2}]_{ij} \cdot F_j 
 \qquad {\rm with} \qquad T_a = \sqrt{\frac{n_F}{6 N_F}} {\rm diag}\{-1,-1,-1, 3/n_F, \cdots, 3/n_F  \}
 \,. 
\end{align}
It is convenient to decompose $F$ into the QCD-charged triplet $(Q)$ and non-charged ones $(\chi)$, so that the $U(1)_{F_A}$ and $U(1)_{PQ}$ transformations above take the form 
\begin{align} 
 F_{i} &= \left(
\begin{array}{c}
 Q \\ 
 \chi 
\end{array}
 \right) 
 \to 
 \left(
\begin{array}{c}
 e^{- i\gamma_5 \alpha_{F_A}/2} Q \\ 
 e^{- i\gamma_5 \alpha_{F_A}/2} \chi 
\end{array}
 \right) 
\qquad {\rm with} \qquad \alpha_{F_A} = \sqrt{\frac{1}{2 N_F}} \alpha_0
\,, \notag\\ 
  F_{i} &= \left(
\begin{array}{c}
 Q \\ 
 \chi 
\end{array}
 \right) 
 \to 
 \left(
\begin{array}{c}
 e^{- i\gamma_5 \alpha_{Q}/2}  Q \\ 
 e^{- i\gamma_5 \alpha_{\chi}/2}  \chi 
\end{array}
 \right) 
\qquad {\rm with} \qquad \alpha_{Q} = - \sqrt{\frac{n_F}{6 N_F}} \alpha_a = - \frac{n_F}{3} \alpha_{\chi}
 \,. 
\end{align}
These anomalous $U(1)_{F_A}$ and $ U(1)_{\rm PQ}$ rotations transform the Lagrangian like 
\begin{align} 
 {\cal L}_{U(1)_{F_A}, U(1)_{\rm PQ}}^{\rm anomalous} 
 \to  
  &- |m_Q|  e^{ - i (\alpha_{F_A} + \alpha_{Q})}  \bar{Q}_{L} Q_{R} 
   - |m_\chi|  e^{ - i ( \alpha_{F_A} + \alpha_{\chi}  )}   \bar{\chi}_{L} \chi_{R}
   +{\rm h.c.}
  \notag\\ 
&  + \frac{\theta_{\rm axicolor} - N_F \alpha_{F_A}}{32 \pi^2} {\cal G}_{\mu\nu} \tilde{\cal G}^{\mu\nu} 
+ \frac{\theta_{\rm QCD} - N_C (\alpha_{Q} + \alpha_{F_A})}{32 \pi^2} {G}_{\mu\nu} \tilde{G}^{\mu\nu} 
\,, 
\end{align}
where $m_Q = m_{1,2,3} =0$ and $m_\chi=m_{4,\cdots,N_F}=m$.

Choosing the  $\alpha_{F_A}$ phase convention, 
\begin{align} 
 \alpha_{F_A} =  \frac{\theta_{\rm axicolor} }{N_F} 
\,, 
\end{align}
we may remove the axicolored topological term. 
Since the axicolor interaction is flavor-blind, 
we require the flavor singlet condition for 
the phase dependent interaction terms, just like the QCD case~\cite{Baluni:1978rf,Kim:1986ax}. 
For infinitesimal phases, we thus have 
\begin{align} 
 |m_Q| (\alpha_{F_A} + \alpha_{Q}) 
 = |m_\chi| ( \alpha_{F_A} + \alpha_{\chi})  
 \equiv x 
\,. 
\end{align}
Note here that $m_Q=0$, hence $x=0$.  
which constrains the PQ and $U(1)_{F_A}$ phases as 
\begin{align} 
  \alpha_{F_A} +\alpha_{\chi} = 0
 \,,  
\end{align}
leaving an arbitrary phase parameter combination, 
\begin{align} 
 \alpha_{F_A} + \alpha_{Q} = {\rm arbitrary} 
 \,. 
\end{align} 
Thus, the axicolored sector is completely CP violating phase-free, 
due to the massless $Q$ fermions, where 
the original $\theta_{\rm axicolor}$ is rotated away by 
the $U(1)_{F_A}$ transformation with $\alpha_{0} = \sqrt{2/N_F} \theta_{\rm axicolor}$ 
and the PQ transformation with $\alpha_a = - \sqrt{2n_F/(3N_F)} \theta_{\rm axicolor}$.

\end{document}